# Field-normalized Impact Factors:
# A Comparison of Rescaling *versus* Fractionally Counted IFs

*Journal of the American Society for Information Science and Technology* (2013, in press)

Loet Leydesdorff,[a]* Filippo Radicchi,[b] Lutz Bornmann,[c] Claudio Castellano,[d] Wouter de Nooy[e]

**Abstract**

Two methods for comparing impact factors and citation rates across fields of science are tested against each other using citations to the 3,705 journals in the *Science Citation Index* 2010 (CD-Rom version of *SCI*) and the 13 field categories used for the *Science and Engineering Indicators* of the US National Science Board. We compare (*i*) normalization by counting citations in proportion to the length of the reference list (1/*N* of references) with (*ii*) rescaling by dividing citation scores by the arithmetic mean of the citation rate of the cluster. Rescaling is analytical and therefore independent of the quality of the attribution to the sets, whereas fractional counting provides an empirical strategy for normalization among sets (by evaluating the between-group variance). By the fairness test of Radicchi & Castellano (2012a), rescaling outperforms fractional counting of citations for reasons that we consider.

**Keywords**: impact, citation, normalization, field, fractional counting, fairness

[a] Amsterdam School of Communication Research (ASCoR), University of Amsterdam, Kloveniersburgwal 48, 1012 CX Amsterdam, The Netherlands; loet@leydesdorff.net; http://www.leydesdorff.net ; *corresponding author.
[b] Universitat Rovira i Virigili, Av. Països Catalans 26, 43007 Tarragona, Spain.
[c] Division for Science and Innovation Studies, Administrative Headquarters of the Max Planck Society, Hofgartenstr. 8, D-80539 Munich, Germany; bornmann@gv.mpg.de.
[d] Istituto dei Sistemi Complessi (ISI-CNR). Via dei Taurini 19, I-00185 Rome, Italy, and Dipartimento di Fisica, Sapienza Università di Roma, P. le A. Moro 2, 00185 Rome, Italy.
[e] Amsterdam School of Communication Research (ASCoR), University of Amsterdam, Kloveniersburgwal 48, 1012 CX Amsterdam, The Netherlands.

**Introduction**

The use of journal impact factors (IFs) for evaluative comparisons across fields of science cannot be justified because fields of science differ in terms of citation practices. In mathematics, for example, reference lists are often short (< 10), while in the bio-sciences reference lists with more than 40 cited references are common. Thus, the *citation potentials* across fields of science are different for purely statistical reasons (Garfield, 1979). Apart from this scale effect, citation distributions have specific characteristics (Albarrán *et al.*, 2011; Glänzel & Schubert, 1988) and thus one may hope to find ways to make them comparable, but only after appropriate normalization. This question of normalization is urgent for the evaluation process because institutional units are rarely monodisciplinary, and thus at the level of institutional units, one can hardly avoid the conundrum of comparing apples with oranges (Rafols *et al.*, 2012). For example, by closing its mathematics department, even if excellent, a university might be able to improve the university's rank in terms of average citation rates.

Small & Sweeney (1985) first proposed using "fractional citation counting," that is, the attribution of citation credit to the cited paper in proportion to the length of the reference list in the citing paper. Zitt & Small (2008) used the audiences of the citing papers as the reference sets for developing Audience Factors of journals—as an alternative to Impact Factors—and Moed (2010) proposed to combine these two ideas when developing SNIP ("Source-Normalized Impact per Paper") as a journal indicator for the Scopus database. Leydesdorff & Opthof (2010) radicalized the idea of fractional counting at the paper level and proposed abandoning normalization in terms of relevant fields that are defined in terms of journal sets, and to use the citing *papers* as the reference sets across fields and journals, and then to attribute citations fractionally from this perspective (cf. Waltman & Van Eck, forthcoming). Using papers as units of analysis also allows for fractional counting of the citations across institutional units with different portfolios (Leydesdorff & Shin, 2011; Zhou & Leydesdorff, 2011). Furthermore, the change to the level of papers for the evaluation allows for statistical decomposition in terms of percentile ranks and hence the use of nonparametric statistics (Bornmann & Mutz, 2011; Leydesdorff *et al.*, 2011)

Returning to journal evaluation, Leydesdorff & Bornmann (2011) decomposed the journal set of the *Science Citation Index* 2008 at the paper level and reconstructed a fractionally counted impact factor. Using numerators from the 3,853 journals included in the CD-Rom version of this database and denominators from the *Journal Citations Report* 2008, these authors found an 81.3% reduction of the between-group variance across 13 major fields distinguished by PatentBoards™ and the US National Science Foundation (NSF) for the biannual evaluation in *Science and Engineering Indicators* (NSB, 2010). The remaining between-group variance was no longer statistically significant. Leydesdorff, Zhou & Bornmann (in press) repeated this analysis using 2010 data, but with more strict criteria, improved statistical methods, and time horizons other than the two-year citation window of the standard impact factor (IF2). As before, the reduction of the between-group variance was 79.2% (as against 81.3% in 2008), but the IF5 further improved on this reduction to 91.7%. The latter result was statistically significant, whereas the former in this case was not.[6]

In the final paragraphs, Leydesdorff, Zhou, & Bornmann (in press) raised the question of how this result would compare to the universal normalization procedure for citation distributions proposed by Radicchi, Fortunato, and Castellano (2008). In this study, we compare the two normalization schemes using the fairness test proposed by Radicchi & Castellano (2012a). This is a statistical test specifically designed for measuring the effectiveness of normalized indicators aimed at the removal of disproportions among fields of science. Radicchi & Castellano (2012a) used this test to show that the rescaled indicator introduced by Radicchi *et al.* (2008) outperformed the fractional indicator proposed by Leydesdorff & Opthof (2010) in the analysis of the citations received by papers published in the journals of the American Physics Society (APS).

Radicchi *et al.*'s (2008) normalization can be applied to any comparison among subsets. The attribution of the cases to the subsets can even be random. The normalized (field-specific) citation count is $c_f = c / c_0$, in which c is the raw citation count and $c_0$ is the average number of citations per unit (article, journal, etc.) for this field—or more generally—this subset. The normalization sets the mean of the scores in each group equal to 1. Consequently, the between-group variance of the rescaled scores is necessarily zero independently of the attribution of the units to the groups.[7]

---

[6] Additionally, the fractionally counted citations using all previous years divided by the number of publications in the current year (2010) or, in other words, a fractionally counted c/p ratio, was not statistically significant across the fields, while it was in the study of Leydesdorff & Bornmann (2011).

Whereas the reasoning of Radicchi and his coauthors (2008, 2012a, 2012b) is analytical and focuses on the homogeneity of the set after normalization, Leydesdorff & Bornmann (2011) studied whether the statistical significance of the dividedness between the groups is reduced by fractionalization as an empirical strategy using the so-called variance components model: in addition to papers being organized in journals (at level 1), journals are at a next level 2 intellectually organized in fields of science. This second-level effect can be measured independently of the first-level effect using multi-level analysis. If the between-group variance is statistically significantly different from zero, the sets' citation impact can be considered as heterogeneous. In other words, the multi-level model (of generalized linear mixed models)

---

[7] The total sum of squares in the analysis of variance ($SS_{tot} = \sum_{j=1}^{J}\sum_{k=1}^{n_j}(X_{jk} - \overline{X})^2$ with $X_{jk}$ = score of case $k$ in subset $j$ and $\overline{X}$ = the mean of the complete set) provides the total spread of the variable $X$. The between-group variance is defined as follows:

$$SS_b = \sum_{j=1}^{J} n_j(\overline{X}_j - \overline{X})^2 \text{ in which formula } \overline{X}_j = \text{the mean of subset } j. \quad (1)$$

When every score $X_{jk}$ is divided by the average for the subset $\overline{X}_j$, each new average of a subset becomes unity since:

$$\overline{X}_j' = \frac{\sum_{k=1}^{n_j}(X_{jk}/\overline{X}_j)}{n_j} = \frac{\sum_{k=1}^{n_j} X_{jk}}{n_j} / \frac{\sum_{k=1}^{n_j}\overline{X}_j}{n_j} = \overline{X}_j / \overline{X}_j = 1 \quad (2)$$

The *grand mean* of the set is equal to the weighted mean of the means of all subsets and therefore also 1:

$$\overline{X}' = \frac{\sum_{j=1}^{J}\sum_{k=1}^{n_j} X_{jk}'}{\sum_{j=1}^{J} n_j} = \frac{\sum_{j=1}^{J}\sum_{k=1}^{n_j} n_j \frac{X_{jk}'}{n_j}}{\sum_{j=1}^{J} n_j} = \frac{\sum_{j=1}^{J} n_j \sum_{k=1}^{n_j} \overline{X}_j'}{\sum_{j=1}^{J} n_j} = \frac{\sum_{j=1}^{J} n_j \overline{X}_j'}{\sum_{j=1}^{J} n_j} = \frac{\sum_{j=1}^{J} n_j 1}{\sum_{j=1}^{J} n_j} = 1 \quad (3)$$

Substitution of Equations 2 and 3 into Equation 1 shows that the between-group variance is zero:

$$SS_b' = \sum_{j=1}^{J} n_j(\overline{X}_j' - \overline{X}')^2 = \sum_{j=1}^{J} n_j(1-1)^2 = 0 \quad (4)$$

*Q.e.d.*

enables us to quantify and statistically test the effects of fractional counting in the comparison among sets, whereas rescaling sets the between-group variance by definition equal to zero.

In this study, we use the same data as in Leydesdorff *et al*. (in press), and compare the fractionally normalized values with the results of normalization based on dividing by the arithmetic mean of the parameter under study (e.g., the IF5) at the level of each cluster, using Radicchi *et al.*'s (2008) rescaling. We rescaled the integer-counted impact factors and their numerators (total citations), and additionally the numerators of IF2 and IF5 as provided by the Journal Citation Reports 2010 of the Science Citation Index-Expanded, but for this same set of 3,705 journals. The project was done in June-August 2012, and at that time the data for 2010 were the most recent data available.

**Methods and materials**

    *a. Data*

Data was harvested from the CD-Rom version of the *Science Citation Index* 2010. This version contains a core set of 3,705 journals contained in the *Science Citation Index-Expanded,* but selected as most representative and used for policy purposes. The U.S. firm PatentBoard™— previously named CHI Inc.—has for several decades been under contract of the U.S. National Science Foundation to add 13 categories to the journal list that is used for the biannual updates of the *Science and Engineering Indicators* of the National Science Board (2012). We use these 13 categories from the 2010 list as the second level, but two categories are not used in the analysis because they are poorly populated in this subset of 3,705 journals: cluster 8 ("Humanities") contained only two journals, and cluster 11 ("Professional fields") only eight journals. Thus, we work with 3,695 journals organized in eleven broad fields of science. The reader is referred to Leydesdorff, Zhou & Bornmann (in press) for further details about the data processing and the distinction of 23 possible variables (including the two- and five-year impact factors).

|     | **Variable** |                                          |
| --- | ------------ | ---------------------------------------- |
| 1.  | ISI-IF2      | IF2 from JCR 2010                        |
| 2.  | ISI-IF5      | IF5 from JCR 2010                        |
| 3.  | IF2-IC       | IF2 integer counted from CD-Rom          |
| 4.  | IF5-IC       | IF5 integer counted from CD-Rom          |
| 5.  | ISI-TC       | Times cited, JCR 2010                    |
| 6.  | TC-IC        | Times cited, integer counted from CD-Rom |
| 7.  | TC-IC2       | IF2 numerator from CD-Rom                |
| 8.  | TC-IC5       | IF5 numerator from CD-Rom                |
| 9.  | IF2-Num      | IF2 numerator from JCR 2010              |
| 10. | IF5-Num      | IF5 numerator from JCR 2010              |
| 11. | c/p 2010     | c/p ratio: variable 5 / variable 13      |

**Table 1**: Variables considered for rescaling. TC=total cites; IC=integer counting; IF=impact factor; JCR=Journal Citation Reports.

Among the 23 variables used by Leydesdorff *et al.* (in press), we use the variables listed in Table 1 for the rescaling procedure in this study. We did not rescale any of the fractionally counted analogues of these integer-counted indicators—IF2-FC, IF5-FC, TC-FC, TC-FC2, and TC-FC5—because the objective of the study is to compare the effects of fractionalization *versus* rescaling as normalization strategies.

Variables 1 and 2—taken from JCR—are different from the corresponding values of variables 3 and 4 because the ISI-IF includes all citations in the larger set of JCR 2010 in the numerator ($N = 10,196$ journals), whereas variables 3 and 4 are based on counting only in the domain of the 3,705 journals included in the CD-Rom version. (The denominators are the same, that is, the sum of citable items in the previous two years as provided by JCR.) The various numerators are separately studied as variables 5 to 10. Finally, variable 11 adds a value derived from JCR: the total cites of each journal (variable 5) divided by the number of this journal's citable items (articles, reviews, and proceedings papers) in the current year 2010.

*b. Methods*

Radicchi & Castellano (2012a) provide a fairness test that can be applied to differently normalized datasets in order to compare whether fractional counting of the citations or rescaling of the citation counts leads to a better result. Note that this is not a trivial question despite the analytical character of rescaling. Different normalizations may have different effects on the distributions of the variables in the various subsets so that variable proportions may belong, for example, to the top-10% of most-highly-cited journals. According to the notion of a fair indicator, the probability of finding a journal with a particular value of this indicator should not depend on the subset of research (e.g., discipline) to which this journal is attributed. The "fairness" of a citation indicator is therefore directly quantifiable by looking at the ability of the indicator to suppress any potential citation bias related to the classification of journals in disciplines or topics of research.

The fairness test was previously used for the assessment of indicators devoted to the suppression of disproportions in citation counts among papers belonging to different sets, but it can straightforwardly be extended in the present case to test the performances of indicators aiming at the suppression of discipline-dependent bias in journal evaluation. In this study, we analyze a set of $N = 3,695$ journals divided into $G=11$ different categories. We indicate with $N_g$ the number of journals within category g. Each journal in the entire set has associated a score calculated according to the rules of the particular indicator we want to test (Table 1). Imagine sorting all journals depending on this indicator and then extracting the top z% of journals from the sorted list. The list of top z% journal is composed of $n^{(z)} = \lfloor zN/100 \rfloor$ journals (where $\lfloor x \rfloor$ indicates the integer part of x).

If the indicator is fair, the presence in the top z% should not depend on the particular category to which the journal belongs. That is, the presence of a journal of category g in the top z% should depend only on $N_g$ and not on the fact that category g is privileged for some other reason. Under these conditions, the number of journals $m_g^{(z)}$ of category g that are part of the top z% is a random variable that obeys the hypergeometric distribution

$$P\left(m_g^{(z)}\middle|n^{(z)}, N, N_g\right) = \binom{N_g}{m_g^{(z)}}\binom{N-N_g}{n^z-m_g^{(z)}} / \binom{N}{n^z} \tag{1}$$

where $\binom{x}{y} = x! / [y! (x-y)!]$ is the binomial coefficient (Radicchi & Castellano, 2012a, at p. 126). By using this distribution, it is therefore possible to estimate the confidence intervals of an ideal fair indicator, and thus one can statistically judge the "fairness" of any other indicator.

**Results**

*a. Rescaling versus fractional counting of the impact factors*

We examined rescaled versions of all the indicators listed in Table 1. Figure 1 shows graphically the outcomes of analyses using the fairness test for the comparison of rescaled values of IF2 and IF5 *versus* their fractionally counted equivalents. In the left column of the first row, the deviations from the 10% expectation are shown for the rescaling of IF2-s and in the right column for fractionally counted IFs-2. The second row repeats the analysis for the case of five-year IFs. Vertically, the graphs are somewhat similar, but horizontally the differences are considerable.

Rescaling outperforms fractional counting: both the summed and average deviances from the 10% score, as well as the standard deviations, are smaller in the case of rescaling (Table 2). Furthermore, the rescaled values passed the test of the 90% confidence interval (assuming a hypergeometric distribution) while the fractionally counted values did not. Thus, the differences in the distributions among scientific fields are effectively removed when one uses the rescaled versions of these indicators.

| Cluster | IF2 Rescaled | IF5 Rescaled | IF2 Fractionally counted | IF5 Fractionally counted |
|---|---:|---:|---:|---:|
| 1. Biology | 9.46 | 9.25 | 5.57 | 5.78 |
| 2. Biomedical Research | 11.35 | 11.15 | 17.70 | 16.73 |
| 3. Chemistry | 10.78 | 11.11 | 11.75 | 12.70 |
| 4. Clinical Medicine | 9.68 | 9.77 | 12.33 | 11.55 |
| 5. Earth & Space | 6.27 | 5.54 | 7.01 | 7.01 |
| 6. Engineering & Tech | 6.53 | 7.88 | 3.15 | 4.27 |
| 7. Health Sciences | 12.50 | 12.50 | 9.38 | 9.38 |
| 9. Mathematics | 17.92 | 15.03 | 4.05 | 5.20 |
| 10. Physics | 11.93 | 12.76 | 10.61 | 11.43 |
| 12. Psychology | 19.05 | 16.67 | 9.52 | 11.90 |
| 13. Social Sciences | 9.68 | 9.68 | 0.00 | 0.00 |
| Mean (± st.dev.) | 11.38 (± 4.03) | 11.03 (± 3.16) | 8.28 (± 4.97) | 8.72 (± 4.75) |
| $\Sigma_i \| x_i - 10 \|$ | 31.91 | 27.11 | 43.72 | 42.68 |

**Table 2**: Percentages of journals belonging to the top-10% set under the different conditions.

Table 2 provides the percentages of journals in the top-decile corresponding to the four panels of Figure 1. If the measure were perfect, the mean would be 10% with a standard deviation depending on the number of papers. Additionally, the sum of the deviations from 10% is provided as another measure. One can conclude that rescaling outperforms fractional counting on this test; IF5 outperforms IF2 using both rescaling and fractional counting. However, this outperformance may also be a statistical fluctuation from the expected value of ten—as we shall see below in Figure 2 and Table 3.

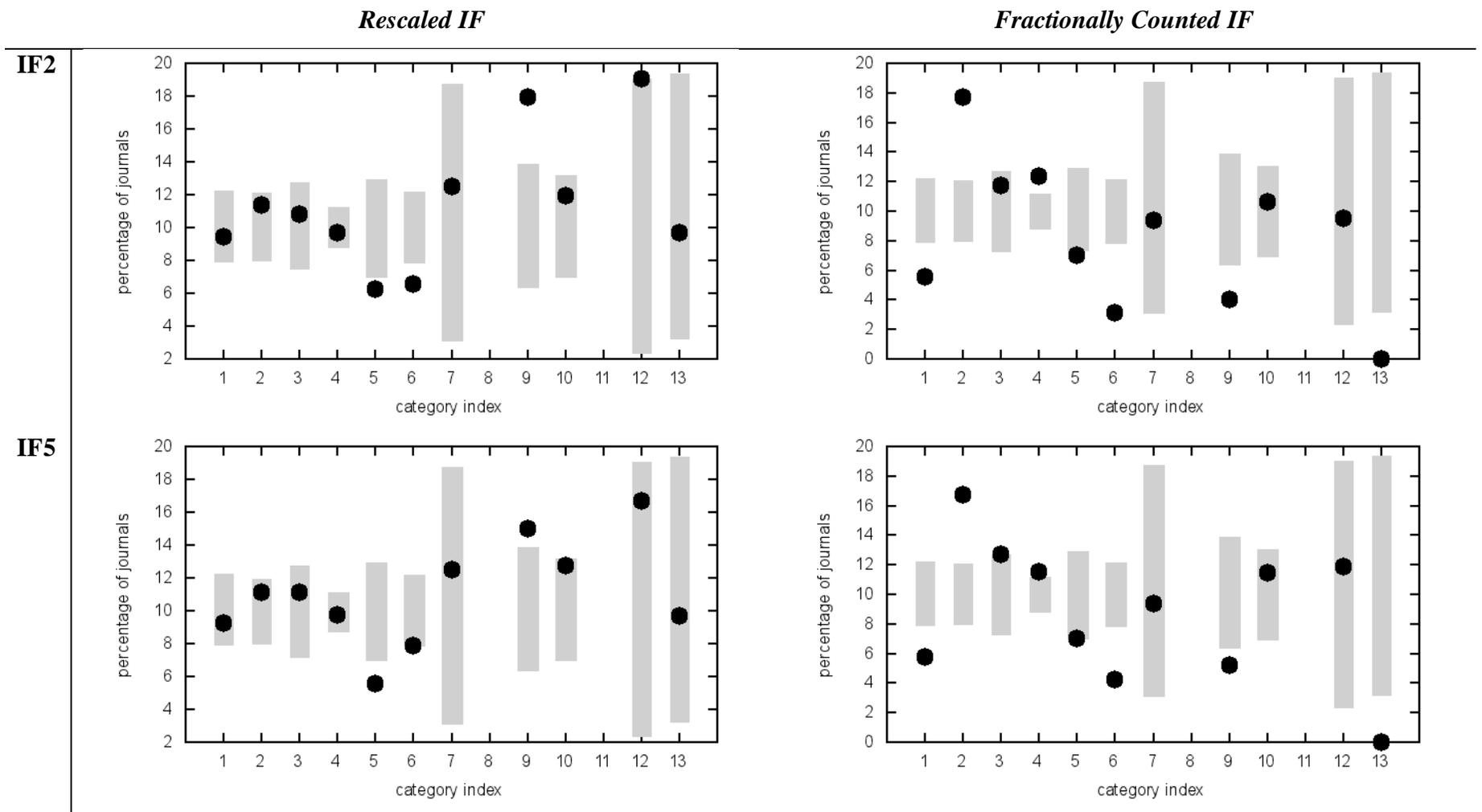

**Figure 1**: Percentages of journals belonging to the top-10% of 3,695 IFs 2010 in eleven different groups, normalized by rescaling and fractional counting of the citations, respectively. Grey areas bound the 90% confidence intervals and are calculated using Eq. (1).

At the level of the individual clusters, fractional counting completely fails the fairness test (with a success rate of 0%) in the case of cluster 13, that is, the "Social Sciences." These are 31 journals attributed to the social sciences, within the domain of the Science Citation Index (and not the Social Science Citation Index). The highest-ranked journal in this deviant set is the *Journal of Human Evolution* which ranks at the 673$^{rd}$ position with ISI-IF2 = 3.843 or 579$^{th}$ position with ISI-IF5 = 4.290. The corresponding ranks are 713$^{th}$ and 556$^{th}$ in the more restricted SCI set under study. Fractionally counted, however, these rankings are worsened to the 726$^{th}$ and 600$^{th}$ positions, respectively. All these values are far outside the domain of the top-10% group of 370 journals ($N = 3695/10 = 369.5$). In the social sciences, referencing is relatively high and citation low in comparison with the natural and life sciences so that fractional counting cannot be expected to improve on the relative standing of these journals in the rankings. By using the arithmetic means of the group as the reference points—the mean values are 0.576 (IF2) and 0.721 (IF5), respectively—rescaling of this set of 31 journals provides *Journal of Human Evolution* with the 127$^{th}$ and 116$^{th}$ positions, respectively, within the set of top-10% highest-ranked journals. However, the number of observations is small in this case.

For another example, let us turn to cluster 9, which is composed of 173 journals designated "Mathematics". Mathematics is the well-known exception in terms of exceptionally low referencing behavior. Might this explain the low value of 5.20% of these journals among the top-10% when using a fractionally counted IF5? The highest IF5 in this group is attributed to *Siam Review* with a value of 3.373. This value ranks the journal at the 428$^{th}$ position and therefore outside the domain of the top-10% of 369 most-highly-ranked journals.[8] Fractionally counted, however, the IF5 of *Siam Review* is upgraded to the 179$^{th}$ position. Three other journals in this group (*Annals of Mathematics* – 115$^{th}$ position; *J American Mathematics Society* – 133$^{rd}$ position; *Acta Mathematica [Djursholm]* – 135$^{th}$ position) are ranked higher than *Siam Review* after fractional counting, among nine journals in total belonging to the top-10% group. Thus, fractional counting in this case corrects for between-field differences. Rescaling brings the fairness test to a value of 15.03%, that is, rather far at the opposite side of the reference standard of ten percent. In the case of "Physics" (cluster 10 with 245 journals), the correction of fractional counting even outperforms rescaling; but this is the exception rather than the rule.

---

[8] The ISI-IF5 of this journal is 5.73; this leads to the 325$^{th}$ position in the ranking, i.e., within the top-10%. (See the discussion about Table 3 below).

Further statistical analysis taught us that the arithmetic means of the fractionally counted citations per cluster correlate significantly with the results of the corresponding parameters on the fairness test ($r = .91$, $p < .01$ for IF2; $r = .92$, $p < .01$ for IF5). This indicates that the fractionally counted IFs still reflect between-field differences. Furthermore, the normalization in terms of fractional counting has uncontrolled effects on the shape of the distributions in terms of standard deviations, skeweness, and kurtosis when comparing across the clusters, whereas rescaling (as a linear transformation) behaves neutrally in this respect.

*b. Rescaling ISI-IFs*

Whereas for the construction of fractionally counted IFs, Leydesdorff, Zhou & Bornmann (in press) needed individual journal-journal citations and where therefore limited to the set of 3,695 journals contained in the CD-Rom/DVD version of the *Science Citation Index* 2010, rescaling can be applied to any set. For Figure 2 and Table 3, we use the same 3,695 journals for comparing ISI-IFs (both for two and five years) in the left columns with the same values divided by the arithmetic means of each of these 11 subsets. Note that in Table 3 the rescaled values of ISI-IF2 outperform the normalization when compared with the rescaled values of ISI-IF5.

| Cluster | ISI-IF2 | ISI-IF5 | ISI-IF2 Rescaled | ISI-IF5 Rescaled |
|---|---|---|---|---|
| 1. Biology | 4.50 | 6.64 | 9.46 | 9.03 |
| 2. Biomedical Research | 20.62 | 19.46 | 12.33 | 12.52 |
| 3. Chemistry | 10.79 | 10.79 | 11.44 | 11.76 |
| 4. Clinical Medicine | 13.53 | 12.41 | 9.33 | 8.90 |
| 5. Earth & Space | 4.80 | 6.64 | 8.49 | 8.86 |
| 6. Engineering & Tech | 2.47 | 3.37 | 9.23 | 9.68 |
| 7. Health Sciences | 9.38 | 12.50 | 12.50 | 12.50 |
| 9. Mathematics | 0.58 | 0.58 | 10.98 | 12.72 |
| 10. Physics | 6.94 | 6.53 | 8.64 | 8.23 |
| 12. Psychology | 16.67 | 19.05 | 11.90 | 11.90 |
| 13. Social Sciences | 0.00 | 0.00 | 16.13 | 16.13 |
| Mean (± st.dev.) | 8.21 (± 6.69) | 8.91 (± 6.62) | 10.95 (± 2.27) | 11.11 (± 2.39) |
| $\Sigma_i \| x_i - 10 \|$ | 62.96 | 60.45 | 20.12 | 22.83 |

**Table 3**: Percentages of journals belonging to the top-10% set when comparing the ISI-IFs of 3,695 journals with their rescaled equivalents.

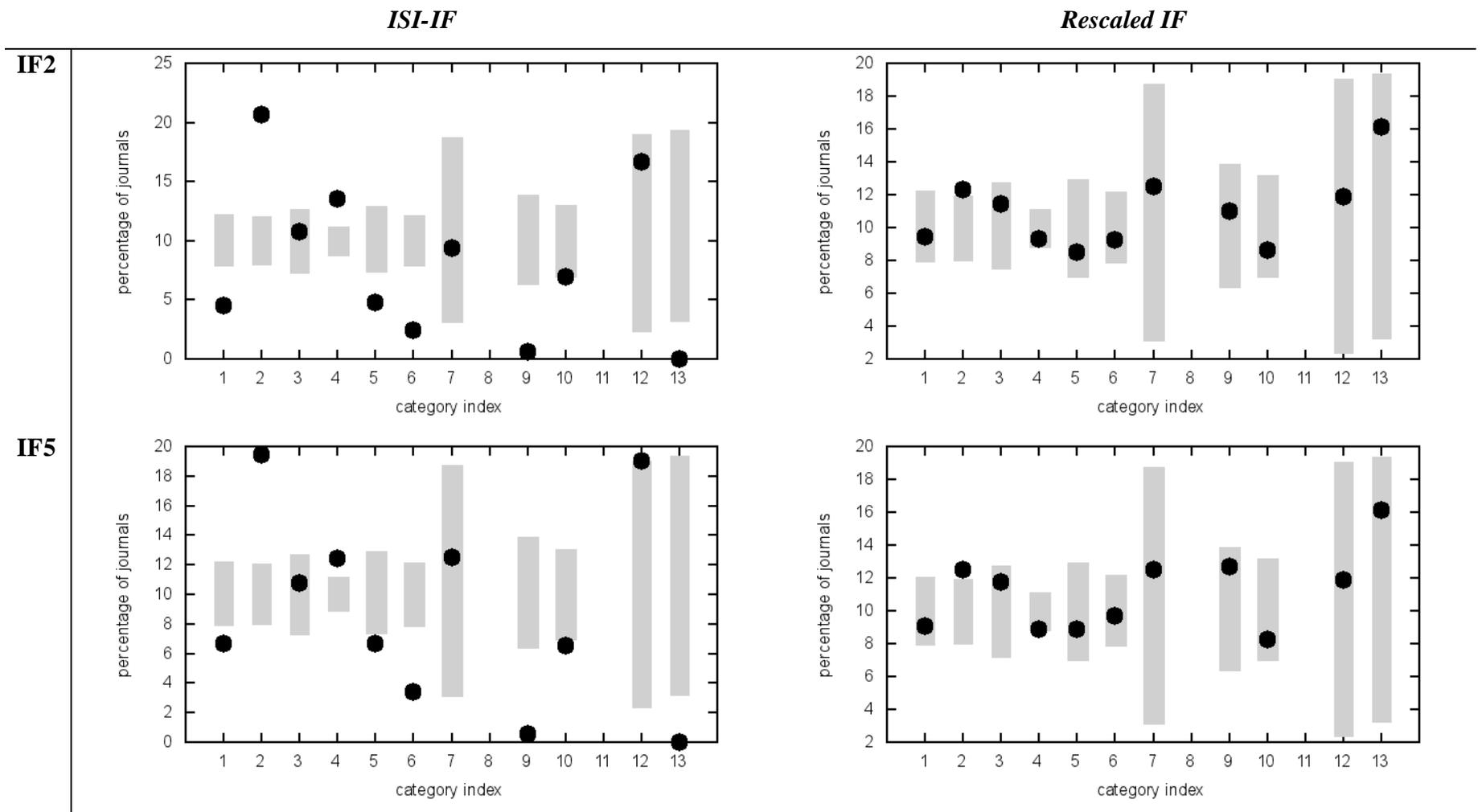

**Figure 2**: Percentages of journals belonging to the top-10% of 3,695 ISI-IFs 2010 in eleven different groups, both non-normalized and normalized by rescaling. Grey areas bound the 90% confidence intervals and are calculated using Eq. (1).

As in the previous comparison, Cluster 13 ("Social Sciences") is not included in the top-10% set when using either ISI-IF2 or ISI-IF5, and only the journal *Siam Review* is within this domain among the 173 mathematics journals (0.58%). Using rescaling, however, the percentages in Table 3 can meaningfully be compared with the reference standard of ten percent.

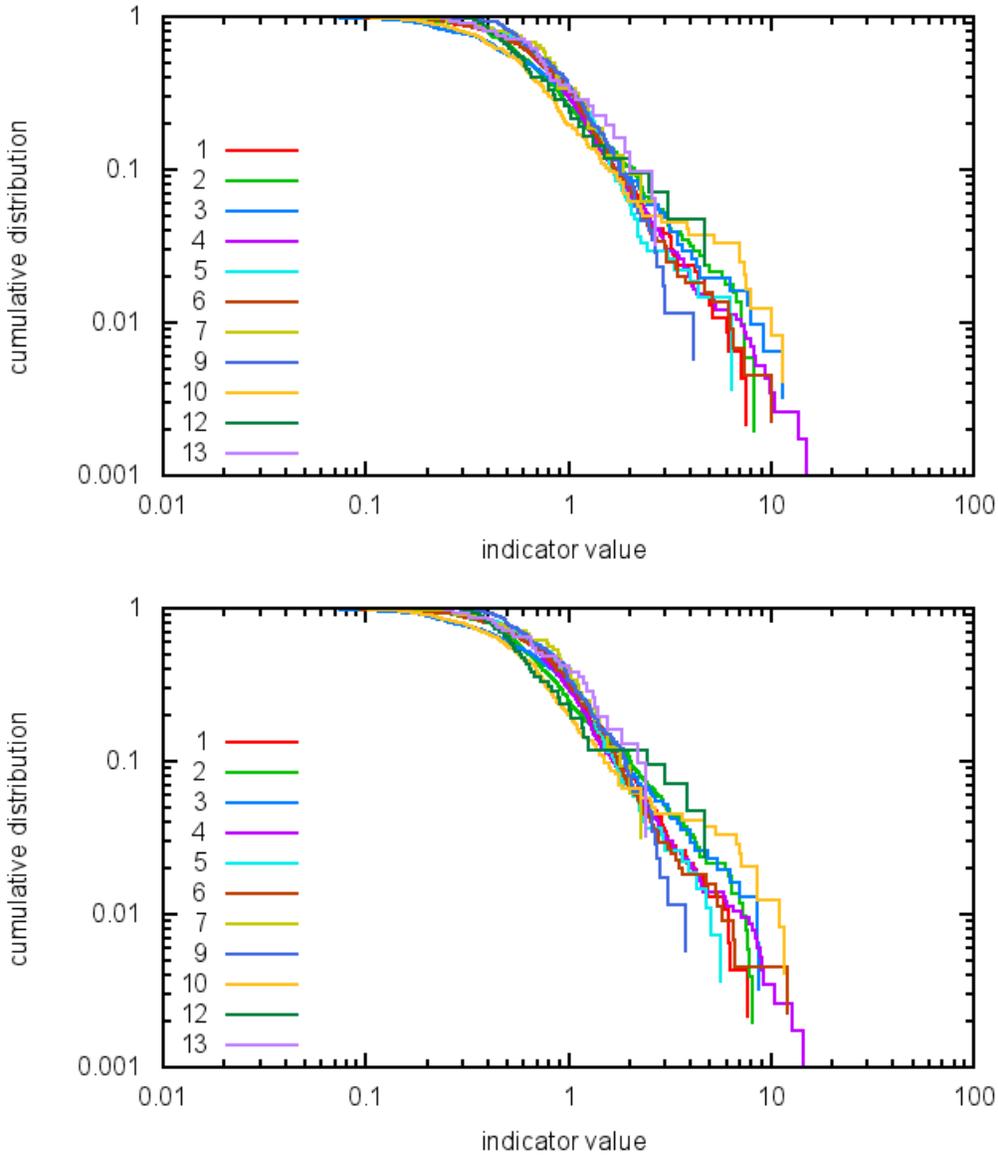

**Figure 3**: Cumulative distribution of rescaled ISI-IFs for eleven groups among 3,695 journals; for IF2 in the top panel and IF5 at the bottom. In this plot, a perfectly fair indicator would have produced a precise collapse of the various curves.

Figure 3 shows the cumulative distributions of the rescaled ISI-IFs as in the right column of Figure 2. The distributions have a similar shape. Differences are small but the curves do not coincide perfectly. Hence the universality that has been claimed for the distributions of article citations within fields (Radicchi *et al*., 2008) is valid only approximately when journal impact factors are considered as citation scores.

**Normalized Impact Factors**

Since the tests indicate that the rescaled impact factors can be compared across fields of science, one can proceed with the comparison. Table 4 lists the top-25 thus normalized ISI-IFs 2010 sorted on the rescaled values of ISI-IF2.

| Rank | Abbreviated journal title | Rescaled ISI-IF2 2010 | Rescaled ISI-IF5 2010 |
|---|---|---|---|
| 1 | CA-CANCER J CLIN | 26.211 | 19.283 |
| 2 | REV MOD PHYS | 19.514 | 18.155 |
| 3 | ACTA CRYSTALLOGR A | 18.881 | 8.576 |
| 4 | NAT MATER | 17.979 | 16.951 |
| 5 | NEW ENGL J MED | 14.872 | 14.380 |
| 6 | ANNU REV PLANT BIOL | 14.063 | 12.002 |
| 7 | ANNU REV IMMUNOL | 13.700 | 12.822 |
| 8 | CHEM REV | 11.479 | 12.641 |
| 9 | ANNU REV ASTRON ASTR | 11.452 | 10.508 |
| 10 | NAT NANOTECHNOL | 11.440 | 11.684 |
| 11 | NAT REV CANCER | 10.338 | 10.403 |
| 12 | NAT PHOTONICS | 9.982 | 11.070 |
| 13 | PROG MATER SCI | 9.970 | 12.036 |
| 14 | NAT REV IMMUNOL | 9.787 | 9.240 |
| 15 | LANCET | 9.352 | 8.925 |
| 16 | CHEM SOC REV | 9.238 | 8.550 |
| 17 | NAT REV MOL CELL BIO | 8.787 | 9.141 |
| 18 | JAMA-J AM MED ASSOC | 8.345 | 8.049 |
| 19 | NAT GENET | 8.270 | 7.190 |
| 20 | NATURE | 8.207 | 7.748 |
| 21 | NAT REV NEUROSCI | 8.206 | 8.995 |
| 22 | ADV PHYS | 8.008 | 7.119 |
| 23 | NAT REV DRUG DISCOV | 7.984 | 8.491 |
| 24 | PROG POLYM SCI | 7.947 | 8.743 |
| 25 | ACCOUNTS CHEM RES | 7.590 | 7.052 |

**Table 4**: 25 journals (abbreviated titles) ordered in terms of the ISI-IF2 after rescaling.

Thus normalized, *Science*, for example, follows only at 36$^{th}$ position with a rescaled IF2 = 7.130 and IF5 = 6.985, whereas it held the 16$^{th}$ position (ISI-IF2 = 31.364) behind *Nature* at the 9$^{th}$ (ISI-IF2 = 36.101) in the JCR 2010. Using fractional counting, *Science* would rank at the 16$^{th}$ (IF2fc = 2.696) and *Nature* at the 13$^{th}$ position (IF2fc = 2.888). However, it should be noted that for the rescaling based on the classification of PatentBoard/NSF, these two journals are not considered as "multidisciplinary science," but as two of 514 journals in the cluster "Biomedical Research," and accordingly rescaled using the arithmetic mean of this subset as denominator. In the case of fractional counting, the attribution to predefined disciplinary groups does not play a role in the normalization because fractional counting is performed at the level of papers and across groups.

The full sets of both rescaled and fractionally counted impact factors 2010 are available online at http://www.leydesdorff.net/if2010/index.htm and at http://www.leydesdorff.net/if2010/normalized_ifs_2010.xlsx , respectively.

**The effects**

A next question is the effect of the various forms of normalization on the overall distributions and rankings, and whether this effect is concentrated specifically in the lower, middle or higher ranks. Figure 3 (above) suggests a different effect for the lowest and highest values of the rescaled indicator. Table 5 first provides the results of a correlation analysis among the different indicators for the comparable sets of 3,695 journals (11 groups).

|              | ISI-IF2 | Rescaled IF2 | Fractionally counted IF2 | ISI-IF5 | Rescaled IF5 | Fractionally counted IF5 |
|---|---|---|---|---|---|---|
| ISI-IF2      |         | .859         | .857                     | .973    | .835         | .815                     |
| Rescaled IF2 | .935    |              | .778                     | .860    | .972         | .763                     |
| Fraction. IF2| .933    | .909         |                          | .826    | .834         | .958                     |
| ISI-IF5      | .977    | .919         | .922                     |         | .883         | .824                     |
| Rescaled IF5 | .913    | .976         | .941                     | .941    |              | .778                     |
| Fraction. IF5| .906    | .896         | .973                     | .932    | .896         |                          |

**Table 5**: Spearman's rank-order correlations $\rho$ organized in the upper triangle and Pearson correlations $r$ in the lower triangle. The *N* of journals varies between 3,675 (for the rescaled values) and 3,695 because of missing values. All correlations are statistically significant at the .01 level.

Table 5 shows that the rescaled IF2 and IF5 correlate across the file precisely as high with each other (Spearman's $\rho = 0.97$ and Pearson's $r = 0.98$) as the unscaled ISI-IF2 and ISI-IF5.[9] The ISI-IFs correlate slightly less with the corresponding normalized IFs, but the rank-order correlations between rescaled and fractionally counted IFs-2 and IFs-5 are only 0.76 and 0.78, respectively. For details about the correlations between fractionally and integer-counted impact factors, the reader is further referred to Leydesdorff, Zhou, and Bornmann (in press: Table 3).

---

[9] When the sets of journals are equal, one would expect the Pearson correlations between IF2 and IF5 to be the same for the original and rescaled IFs because rescaling extracts the between groups variation both from the numerator (covariance between the two variables) and from the denominator (product of the standard deviations of the two variables). If so, the equality among the correlations is analytical. In our case, however, the numbers of journals are different because they were taken in the one case from the Web of Science and in the other from the CD-Rom version.

Decomposition of the correlations into deciles shows that the rescaled values and fractionally counted values of quasi-IFs correlate highest with ISI-IFs in the top and bottom deciles. Figure 4 provides the graphs for IF2. This figure is almost similar for IF5 (not shown here).

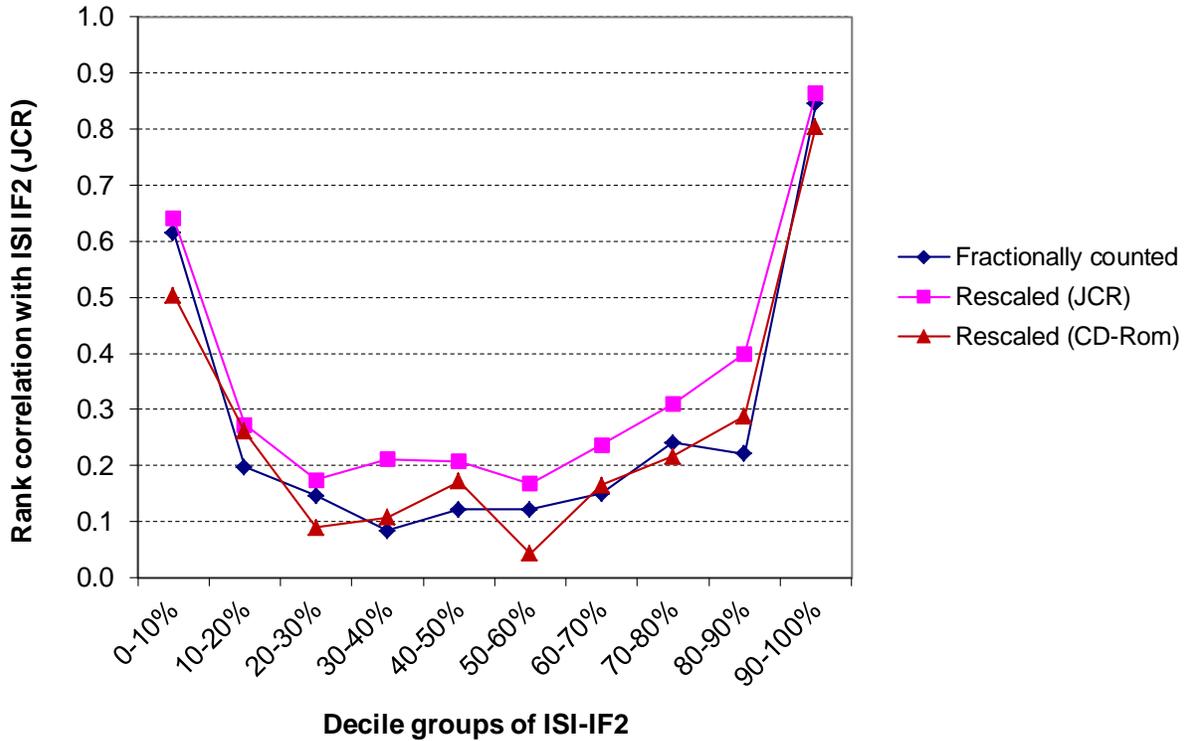

**Figure 4**: Rank-order correlations between ISI-IF2 and normalized values based on scaling the JCR values (■), scaling on the basis of the CD-Rom version (▲), and fractional counting (♦).

This decomposition informs us that the high correlations between the original ISI-IFs and the normalized ones at the aggregated level are an effect of the top and bottom percentiles. Figure 5 shows the scatterplots of rescaled *versus* original IFs-2, in the top row for the lowest two deciles and in the bottom row for the highest two.

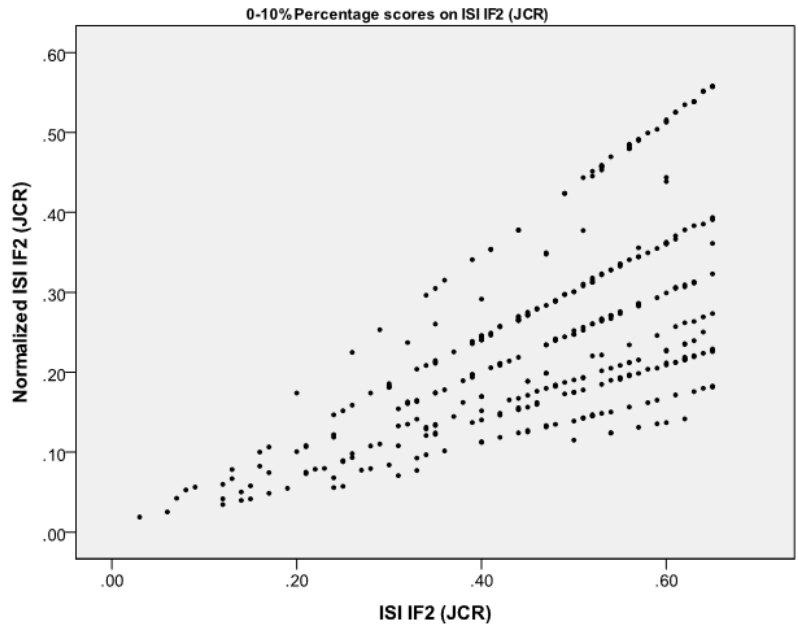
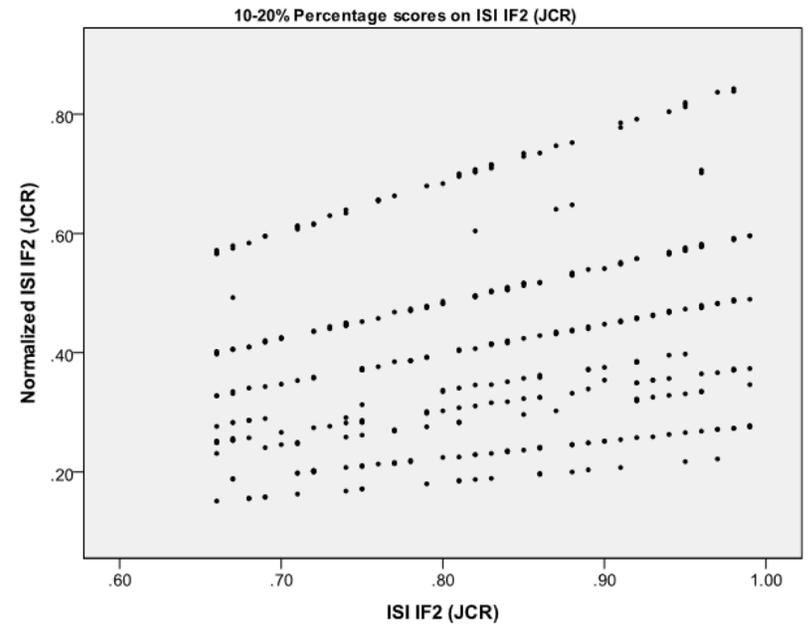
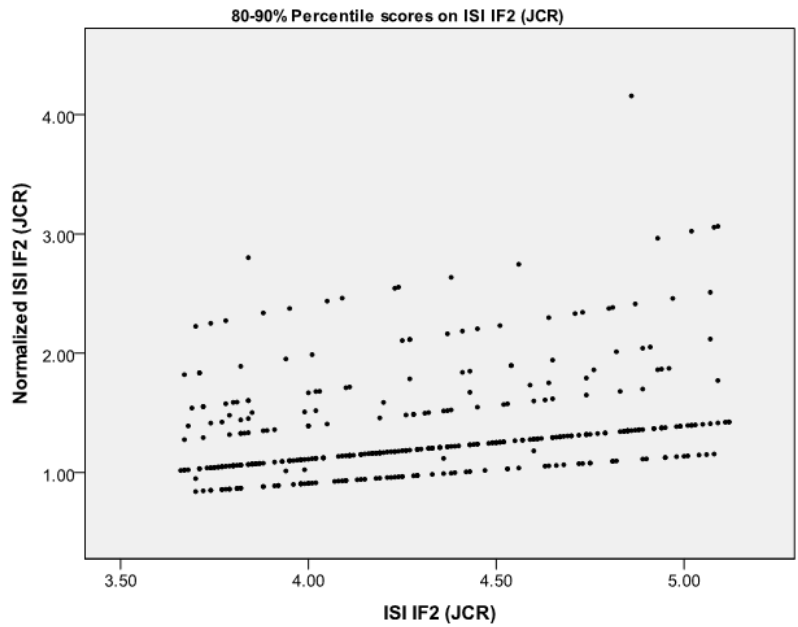
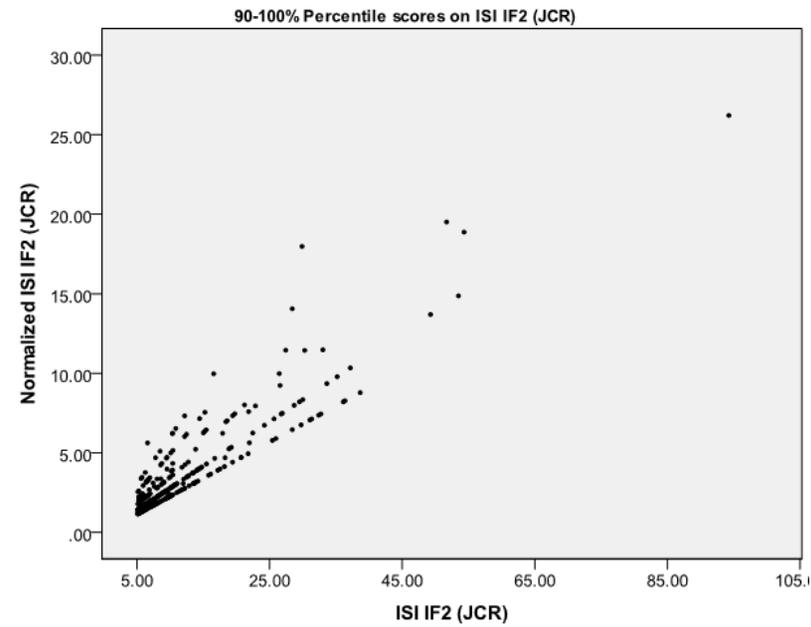

**Figure 5**: Scattergrams of rescaled *versus* original ISI-IF2 for four deciles.

The second and the ninth deciles show a weak correlation as do the deciles with in-between numbers. In the first decile, rescaled values (or equivalently fractionally counted values) can hardly differ from the original values because journals with very few or no citations must occupy the lowest ranks in any ranking (before or after normalization); this correlation is thus high for technical reasons. Indeed, the IFs are indeed clustered in a relatively small strip in the left part of this scattergram. Once the numbers of citations allow for more substantial variation among the normalization outcomes, in the right part of this scattergram, the observations fan out, leading to a somewhat lower correlation.

In the tenth decile, finally, some outliers with high IFs generate a correlation relative to larger clusters at the bottom left. In other words, the strong correlation among the indicators in the top-10% group of journals indicates that normalization is not needed in *this* subset. The top-10% has also been recommended as a relatively robust "excellence indicator" in other contexts (Bornmann & Leydesdorff, 2011; Bornmann *et al.*, 2012; Leydesdorff & Bornmann, 2012; Waltman *et al.*, 2012). The skewness of the distribution makes this segment relatively insensitive to normalization by rescaling or fractioning of the citations. The rankings among the vast majority of journals are, however, very sensitive to different normalizations for between-field differences because the differences among the IFs of journals are often minute, to the second or third decimal.

**Conclusions and Discussion**

In this study, our two teams joined forces to address the question raised by Radicchi & Castellano (2012a) about comparing the fractional counting of citations with rescaling by dividing by the arithmetic mean of each subset, using the complete set of journals studied by Leydesdorff, Zhou & Bornmann (in press) to generate quasi-IFs. The original idea was to apply the multi-level method used in the latter study also to the set of rescaled values so that the variance components could be specified and made comparable. However, rescaling annihilates between-group variance because all the arithmetic means of the groups are set at unity. The two sets of values could therefore not be compared using this method.

Radicchi & Castellano (2012a) proposed a "fairness test" that was applied to APS publications and showed that rescaling outperformed fractional counting in this case. Our results confirm this conclusion. The fairness test was even more convincing when applied to the ISI-IFs provided by the *JCR 2010* of the Web of Science (based on 10,196 journals) then to the integer-counted citations to 3,695 journals which provided the basis for our study of fractionally counted citations. However, the correlation in the top-10% among non-normalized and (differently) normalized values of IFs is high (Figure 4).

Rescaling makes it possible to compare across differently grouped sets because the resulting distributions are, at least approximately, "universal" (Figure 3). The distributions are highly comparable (at least within this set of journals; cf. Waltman *et al.* [2012]). The law of cumulative advantages as specified by Price (1976) or other mechanisms dictating the shape of citation distributions thus seem to operate field-independently; that is, the log-log distribution remains after correction for the differences among fields by using rescaling. At the top- and bottom-ends of the distributions, however, considerable deviance from this "universal" regularity is also visible (Leydesdorff & Bensman, 2006).

The different objective of the multi-level approach remains that one can specify the reduction of between-group variance and test the remaining between-group variance on its deviance from zero. In other words, rescaling is insensitive to the quality of the clustering, whereas the variance decomposition based on fractional counting can also be quantified among alternative groupings. Fractional counting can further be improved (and tested!) using methods recently specified by Waltman & Van Eck (forthcoming).

In this study, the different forms of normalizations were applied to journal impact factors (Garfield, 1972). Criticism of this measure for the evaluation of journals (e.g., Seglen, 1997) and *a fortiori* for the evaluation of papers within journals should in this context be mentioned (Braun, 2007; Lonzano *et al.*, 2012; Leydesdorff, 2012; Vanclay, 2012). More recently, however, book citations (Kousha *et al.*, 2011; Leydesdorff & Felt, 2012) have been added to the potential candidates for impact evaluation. The reasoning here above is not confined to journal evaluation.

When one compares across heterogeneous sets—for example, in the case of evaluating composed sets such as universities with departments and/or when it is difficult to distinguish crisp sets—one can be advised to use rescaling because the quality of the attribution of cases to clusters cannot invalidate this method. Note that one can rescale any variable that differs systematically across sets (e.g., publication rates). One pragmatic advantage in the case of citations, however, is that citation analysis of the citing papers is not needed before rescaling, while the full audience set is required for computation in the case of fractional counting.

**Acknowledgment**

We thank Thomson-Reuters for access to the data. We thank Kim Hamilton for providing the journal list of Patent-Board, and NSF for giving permission.